\documentclass[twocolumn]{ceurart}

\begin{document}

\copyrightyear{2021}
\copyrightclause{Copyright for this paper by its authors.
  Use permitted under Creative Commons License Attribution 4.0
  International (CC BY 4.0).}

\conference{DISCO 2021: Digital Infrastructures for Scholarly Content Objects, September 30--October 01, 2021, Online}

\title{An Open-Publishing Response to the COVID-19 Infodemic}

  \author[1,2,3]{Halie M. Rando}[%
			orcid=0000-0001-7688-1770,
				email=halie.rando@cuanschutz.edu
	]
			\address[1]{
							University of Colorado School of Medicine,
										Center for Health AI,
										Aurora,
										CO,
										USA
					}
			\address[2]{
							University of Colorado School of Medicine,
										Department of Biochemistry and Molecular Genetics,
										Aurora,
										CO,
										USA
					}
			\address[3]{
							University of Pennsylvania,
										Perelman School of Medicine, Department of Systems Pharmacology and Translational Therapeutics,
										Philadelphia,
										PA,
										USA
					}
	  \author[4]{Simina M. Boca}[%
			orcid=0000-0002-1400-3398,
				email=smb310@georgetown.edu
	]
			\address[4]{
							Georgetown University Medical Center,
										Innovation Center for Biomedical Informatics,
										Washington,
										DC,
										USA
					}
	  \author[5]{Lucy {D'Agostino McGowan}}[%
			orcid=0000-0001-7297-9359,
				email=lucydagostino@gmail.com
	]
			\address[5]{
							Wake Forest University,
										Department of Mathematics and Statistics,
										Winston-Salem,
										NC,
										USA
					}
	  \author[3,6]{Daniel S. Himmelstein}[%
			orcid=0000-0002-3012-7446,
				email=daniel.himmelstein@gmail.com
	]
			\address[6]{
							Related Sciences
																	}
	  \author[7]{Michael P. Robson}[%
			orcid=0000-0002-4859-0033,
				email=michael.robson@villanova.edu
	]
			\address[7]{
							Villanova University,
										Department of Computing Sciences,
										Villanova,
										PA,
										USA
					}
	  \author[1,3]{Vincent Rubinetti}[%
			orcid=0000-0002-4655-3773,
				email=vince.rubinetti@gmail.com
	]
	  \author[8]{Ryan Velazquez}[%
			orcid=0000-0002-3655-3403,
				email=rnhvelazquez@gmail.com
	]
			\address[8]{
							Azimuth1,
													McLean,
										VA,
										USA
					}
	  \author[]{{COVID-19 Review Consortium}}[%
	]
			\address[]{
																	}
	  \author[1,2,3,9]{Casey S. Greene}[%
			orcid=0000-0001-8713-9213,
				email=greenescientist@gmail.com
	]
			\address[9]{
							Alex's Lemonade Stand Foundation,
										Childhood Cancer Data Lab,
										Philadelphia,
										PA,
										USA
					}
	  \author[10,11]{Anthony Gitter}[%
			orcid=0000-0002-5324-9833,
				email=gitter@biostat.wisc.edu
	]
			\address[10]{
							University of Wisconsin-Madison,
										Department of Biostatistics and Medical Informatics,
										Madison,
										WI,
										USA
					}
			\address[11]{
							Morgridge Institute for Research,
													Madison,
										WI,
										USA
					}
	
\hypertarget{abstract}{%
\label{abstract}}
\begin{abstract}
The COVID-19 pandemic catalyzed the rapid dissemination of papers and preprints investigating the disease and its associated virus, SARS-CoV-2.
The multifaceted nature of COVID-19 demands a multidisciplinary approach, but the urgency of the crisis combined with the need for social distancing measures present unique challenges to collaborative science.
We applied a massive online open publishing approach to this problem using Manubot.
Through GitHub, collaborators summarized and critiqued COVID-19 literature, creating a review manuscript.
Manubot automatically compiled citation information for referenced preprints, journal publications, websites, and clinical trials.
Continuous integration workflows retrieved up-to-date data from online sources nightly, regenerating some of the manuscript's figures and statistics.
Manubot rendered the manuscript into PDF, HTML, LaTeX, and DOCX outputs, immediately updating the version available online upon the integration of new content.
Through this effort, we organized over 50 scientists from a range of backgrounds who evaluated over 1,500 sources and developed seven literature reviews.
While many efforts from the computational community have focused on mining COVID-19 literature, our project illustrates the power of open publishing to organize both technical and non-technical scientists to aggregate and disseminate information in response to an evolving crisis.
\end{abstract}

\hypertarget{keywords}{%
\label{keywords}}

\maketitle

\hypertarget{introduction}{%
\section{INTRODUCTION}\label{introduction}}

\begin{figure}[htbp]
\hypertarget{fig:cord19-growth}{%
\centering
\includegraphics[width=0.5\textwidth]{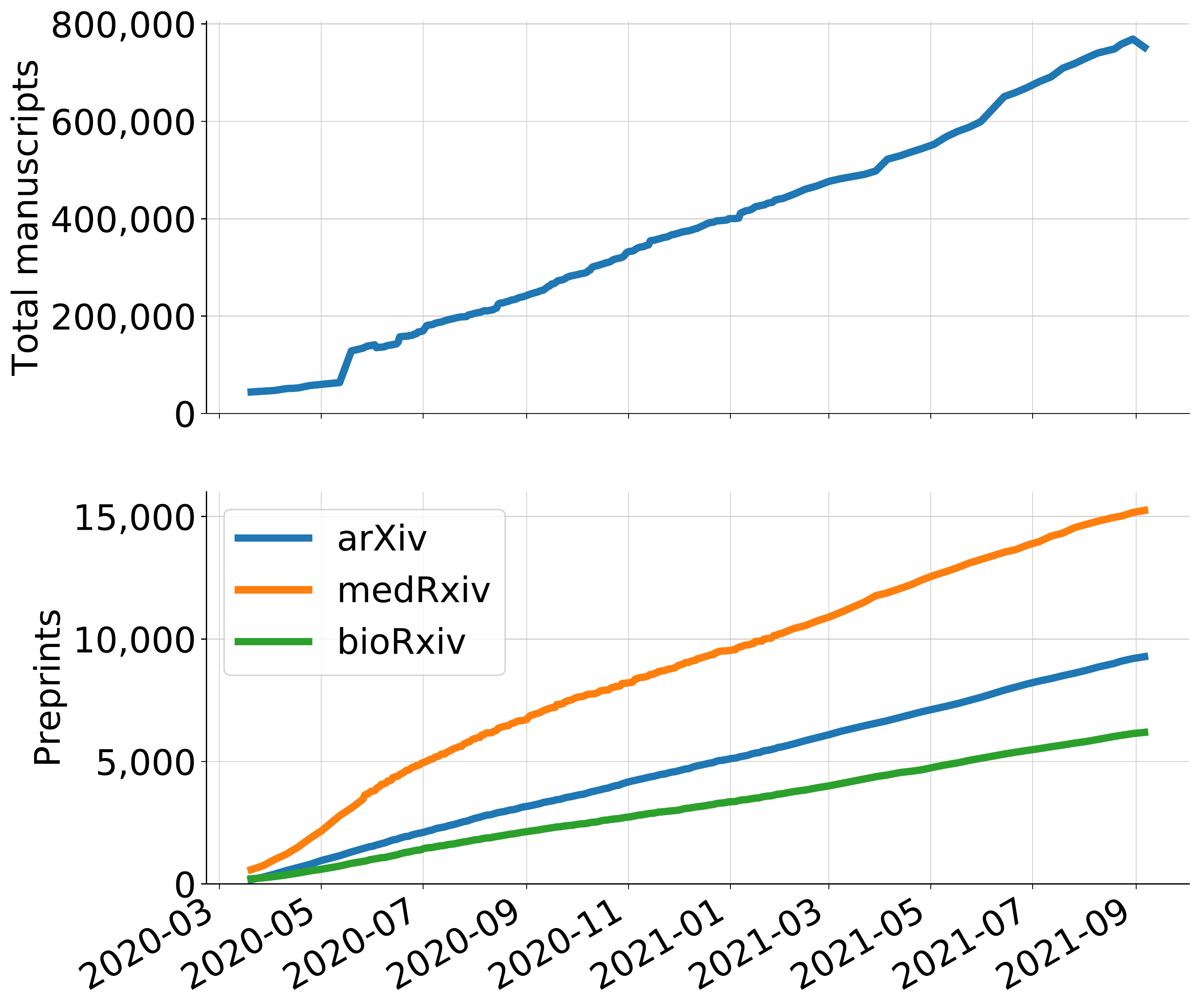}
\caption{\textbf{Growth of the CORD-19 dataset.}
The number of articles has proliferated, with both traditional and preprint manuscripts in the corpus.
The first release (March 16, 2020) contained 28,000 documents \citep{CiOwklc6}.
As of September 6, 2021, this had increased to 768,929 articles.
Of these, 30,726 are preprints from \emph{arXiv}, \emph{medRxiv}, and \emph{bioRxiv}.
}\label{fig:cord19-growth}
}
\end{figure}

Coronavirus Disease 2019 (COVID-19) caused a worldwide public health crisis that has reshaped many aspects of society.
The scientific community has, in turn, devoted significant attention and resources towards COVID-19 and the associated virus, SARS-CoV-2, resulting in the release of data and publications at a rate and scale never previously seen for a single topic.
Over 20,000 articles about COVID-19 were released in the first four months of the pandemic \citep{7ub6VM4Z}, causing an  ``infodemic'' \citep{7ub6VM4Z, nnfOazAC}.
The COVID-19 Open Research Dataset (CORD-19) \citep{CiOwklc6}, which was developed in part with the goal of training machine learning algorithms on COVID-19-related text, illustrates the growth of related scholarly literature (Figure \ref{fig:cord19-growth}).
This resource was developed by querying several sources for terms related to SARS-CoV-2 and COVID-19, as well as the coronaviruses SARS-CoV-1 and MERS-CoV and their associated diseases \citep{CiOwklc6}.
CORD-19 contained 768,929 manuscripts as of September 6, 2021.
Additional curation by CoronaCentral \citep{Ybg667S0} has produced, at present, a set of over 180,000 publications particularly relevant to COVID-19 and closely related viruses.
Despite many advances in understanding the virus and the disease, there are also downsides to the availability of so much information.
"Excessive publication" has been recognized as a concern for over forty years \citep{DfSr1Ohc} and has been discussed with respect to the COVID-19 literature \citep{dUvvJvv6}.
Any effort to synthesize, summarize, and contextualize COVID-19 research will face a vast corpus of potentially relevant material.

Information was released rapidly by both traditional publishers and preprint servers, and many papers faced subsequent scrutiny.
The number of COVID-19 papers retracted may be higher, and potentially much higher, than is typical, although a thorough investigation of this question requires more time to elapse \citep{ZUk10707, caxpZEmy}.
Many preprints and papers are also associated with corrections or expressions of concern\footnote{\url{https://asapbio.org/preprints-and-covid-19} as well as \url{https://retractionwatch.com/retracted-coronavirus-covid-19-papers}} \citep{caxpZEmy}.
Preprints are released prior to peer review, but some traditional publishing venues have fast-tracked COVID-19 papers through peer review, leading to questions about whether they are held to typical standards \citep{1Dez1ZOc5}.
Therefore, evaluating the COVID-19 literature requires not only digesting available information but also monitoring subsequent changes.

Because of the fast-moving nature of the topic, many efforts to summarize and synthesize the COVID-19 literature have been undertaken.
These efforts include newsletters\footnote{\url{https://depts.washington.edu/pandemicalliance/covid-19-literature-report/latest-reports}} \citep{JdWiPJCL}, web portals\footnote{\url{https://outbreaksci.prereview.org}} \citep{1CBWvhTdy} or the now-defunct http://covidpreprints.com\footnote{\url{https://asapbio.org/preprints-and-covid-19}}, comments on preprint servers\footnote{\url{https://disqus.com/by/sinaiimmunologyreviewproject}} \citep{YZ4cHNuH}, and even a journal\footnote{\url{https://rapidreviewscovid19.mitpress.mit.edu}}.
However, the explosive rate of publication presents challenges for such efforts, many of which are no longer active.
Similarly, many literature reviews have been written on the available COVID-19 literature \citep{I2EsJmfs, evtsR3C5, 18eCxyLhx, SAE5ME3N, xOs5ctsW}, but static reviews quickly become outdated as new research is released or existing research is retracted or superseded.
One example is a review of topics in COVID-19 research including vaccine development \citep{xOs5ctsW}.
This review was published on July 10, 2020, four days before Moderna released the surprisingly promising results of their phase 1 trial \citep{wiGjCZC8} that changed expectations surrounding vaccines.
Therefore, the COVID-19 publishing climate presented a challenge where curation of the literature by a diverse group of experts in a format that could respond quickly to high-volume, high-velocity information was desirable.

We therefore sought to develop a platform for scientific discussion and collaboration around COVID-19 by adapting open publishing infrastructure to accommodate the scale of COVID-19 publishing.
Recent advances in open publishing have created an infrastructure that facilitates distributed, version-controlled collaboration on manuscripts \citep{YuJbg3zO}.
Manubot \citep{YuJbg3zO} is a collaborative framework developed to adapt open-source software development techniques and version control for manuscript writing.
With Manubot, manuscripts are managed and maintained using GitHub, a popular, online version control interface.
We selected Manubot because it offers several advantages over comparable collaborative writing platforms such as Authorea, Overleaf, Google Docs, Word Online, or wikis \citep{YuJbg3zO}.
Citation-by-identifier ensures consistent reference metadata standards that would be difficult to maintain manually in a manuscript with dozens of authors and over 1,500 citations.
Manubot's pull request-based contribution model balances the goals of making the project open to everyone and maintaining scientific accuracy.
All contributions are reviewed, discussed, and formally approved on GitHub before text updates appear in the public-facing manuscript\footnote{\url{https://greenelab.github.io/covid19-review}}.
Continuous integration (CI) seamlessly combines author-produced text and figures with automatically generated and updated statistics and figures derived from external data sources and the manuscript's own content.
In addition, the authors who initially launched this project included Manubot developers who had prior successes using Manubot for massively open and traditional manuscript, such as a large-scale collaborative efforts such as a review of developments in deep learning \citep{PZMP42Ak} and a re-evaluation of the role of authorship in modern collaborations \citep{6acsZuy7}.

Collaboration via massively open online papers has been identified as a strategy for promoting inclusion and interdisciplinary thought \citep{PoDz2q0A}.
However, the Manubot workflow can be intimidating to contributors who are not well-versed in git \citep{PoDz2q0A}.
The synthesis and discussion of the emerging literature by biomedical scientists and clinicians is imperative to a robust interpretation of COVID-19 research.
Such efforts in biology often rely on What You See Is What You Get tools such as Google Docs, despite the significant limitations of these platforms in the face of excessive publication.
We recognized that the problem of synthesizing the COVID-19 literature lent itself well to the Manubot platform, but that the potential technical expertise required to work with Manubot presented a barrier to domain experts.

Here, we describe the adaptation of Manubot to facilitate collaboration in the extreme case of the COVID-19 infodemic, with the objective of developing a centralized platform for summarizing and synthesizing a massive amount of preprints, news stories, journal publications, and data.
Unlike prior collaborations built on Manubot, most contributors to the COVID-19 collaborative literature review came from biology or medicine.
The members of the COVID-19 Review Consortium consolidated information about the virus in the context of related viruses and to synthesize rapidly emerging literature.
Manubot provided the infrastructure to manage contributions from the community and create a living, scholarly document integrating data from multiple sources.
Its back-end allowed biomedical scientists to sort and distill informative content out of the overwhelming flood of information \citep{1HZeeO4Cs} in order to provide a resource that would be useful to the broader scientific community.
This case study demonstrates the value of open collaborative writing tools such as Manubot to emerging challenges.
Because it is open source software, we were able to adapt and customize Manubot to flexibly meet the needs of COVID-19 review.
Recording the evolution of information over time and assembling a resource that auto-updated in response to the evolving crisis revealed the particular value that Manubot holds for managing rapid changes in scientific thought.

\hypertarget{methods}{%
\section{METHODS}\label{methods}}

\hypertarget{contributor-recruitment-and-roles}{%
\subsection{Contributor Recruitment and Roles}\label{contributor-recruitment-and-roles}}

First, it was necessary to establish Manubot as a platform accessible to researchers with limited experience working with version control, given that this is not typically emphasized in biology and medicine \citep{1HmO21gZN, OO1DuZd, 4ny1onB0}.
Contributors were recruited primarily by word of mouth and on Twitter, and we also collaborated with existing efforts to train early-career researchers.
We invited potential collaborators to contribute a short introduction on a GitHub issue in order to collect information about participants and provide an introduction to working with GitHub issues.
Interested participants were encouraged to contribute in several ways.
One option was to catalog articles of interest as issues.
We developed a standardized set of questions for contributors to consider when evaluating an article following a framework often used for assessing medical literature.
This approach emphasizes examining the methods used, assignment (whether the study was observational or randomized), assessment, results, interpretation, and how well the study extrapolates \citep{17OQtAY4l}.
Contributors were also invited to contribute or edit text using GitHub's pull request system.
These contributions were not strictly defined and could range from minor corrections to punctuation and grammar to large-scale additions of text.
Finally, a small number of contributors (the authors of this paper) contributed technical expertise, either through the development of standardized approaches to the evaluation of papers based on the MAARIE Framework \citep{k1GJLUxP}, the writing of code to generate manuscript figures, or the addition of features to Manubot.
All of these additions were also submitted as pull requests, either to the COVID-19 review repository or to an external repository, as appropriate.

Each pull request was reviewed and approved by at least one other contributor before being merged into the main branch.
We tagged potential reviewers based on the introductions they had contributed in order to encourage participation.
Authorship was determined based on the Contributor Roles Taxonomy\footnote{\url{https://casrai.org/credit}}.
Due to the permeability of ideas among different sections, contributors to a specific manuscript were recognized with masthead authorship, while all contributors to the project were recognized with consortium authorship on all papers.
Emphasizing the use of issues and pull requests was designed to encourage authors with and without git experience to discuss papers and provide feedback (both formal and informal) on proposed text additions or changes.
We also used the Gitter chat platform\footnote{\url{https://www.gitter.im}} to promote informal questions and sharing of information among collaborators.

\hypertarget{utilization-and-expansion-of-manubot}{%
\subsection{Utilization and Expansion of Manubot}\label{utilization-and-expansion-of-manubot}}

Applying Manubot's existing capabilities allowed us to confront several challenges common in large-scale collaborations, such as maintaining a record of contributions that allowed us to allocate credit appropriately or to contact the original author if questions arose.
Additionally, an up-to-date version of the content was available at all times online in HTML\footnote{\url{https://greenelab.github.io/covid19-review}} or PDF format\footnote{\url{https://greenelab.github.io/covid19-review/manuscript.pdf}}.
This approach also allowed us to minimize the demand on authors to curate and sync bibliographic resources.
Manubot provides the functionality to create a bibliography using digital object identifiers (DOIs), website URLs, or other identifiers such as PubMed identifiers and arXiv IDs.
The author can insert a citation in-line using a format such as \texttt{{[}@doi:10.1371/journal.pcbi.1007128{]}}.
Manubot then obtains reference metadata, exports the citations as Citation Style Language JSON Data Items, and renders the bibliographic information needed to generate the references section \citep{YuJbg3zO}.
This approach allows multiple authors to work on a piece of text without needing to make manual adjustments to the reference lists.

\begin{figure*}[htbp]
\hypertarget{fig:manubot-workflow}{%
\centering
\includegraphics[width=\textwidth]{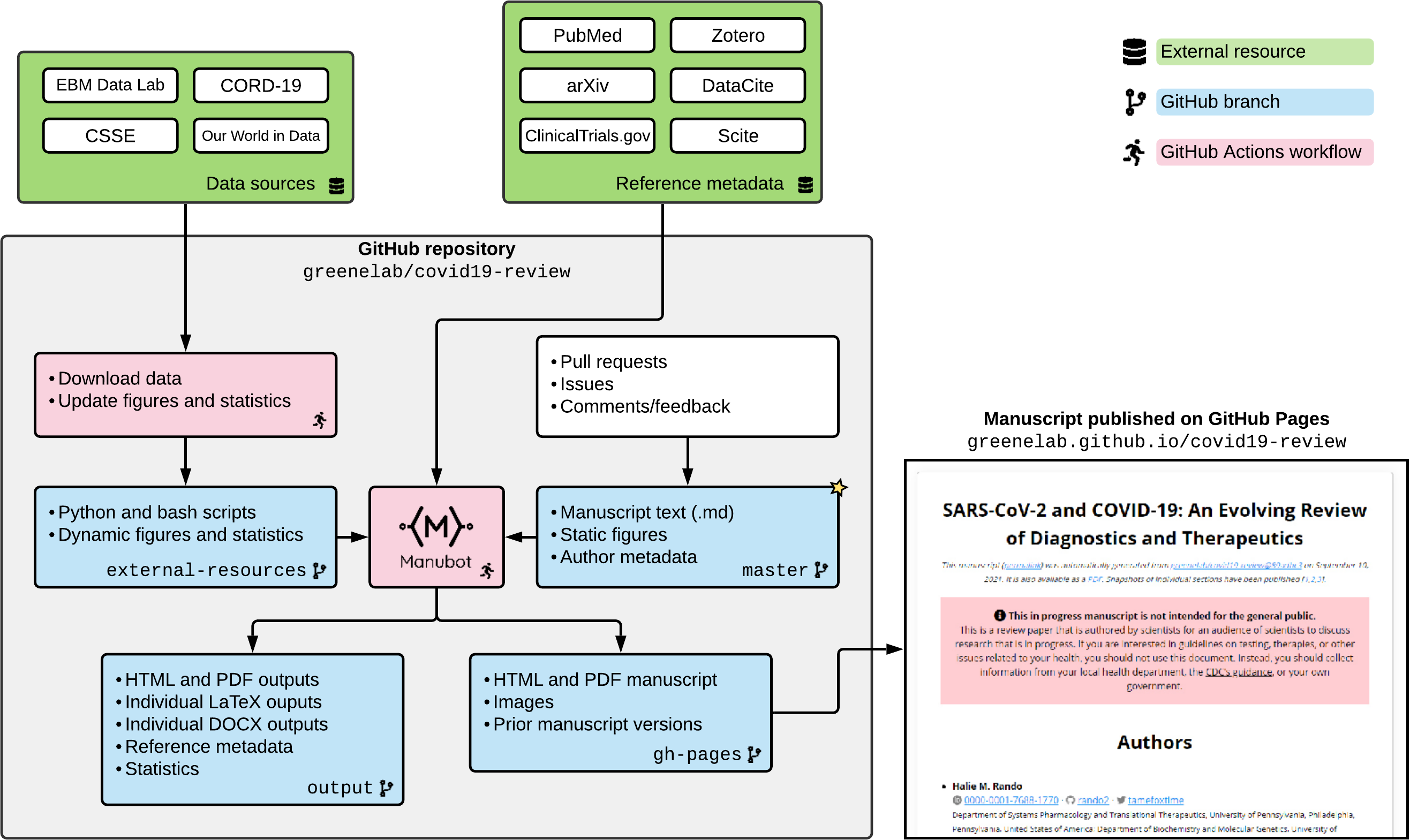}
\caption{\textbf{COVID-19 review GitHub repository organization and workflows.}
Manubot uses CI to combine author-contributed content with automatically updated information from outside sources.
A nightly workflow updates figures and statistics derived from external resources.
Authors write text and add figures to the \texttt{master} branch (starred) via GitHub pull requests.
Manubot generates updated manuscript outputs for each new git commit, integrating the static text and figures with the dynamic statistics and figures and automatically-extracted citation information.
GitHub Pages hosts the latest HTML and PDF versions of the manuscript along with permanent links to prior versions.}\label{fig:manubot-workflow}
}
\end{figure*}

Due to the needs of this project, several new features were implemented in Manubot.
Because of the ever-evolving nature of the COVID-19 crisis, figures and statistics in the text quickly became outdated.
To address this concern, Manubot and GitHub's CI features were used to create figures that integrated online data sources and to dynamically update information, such as the current number of active COVID-19 clinical trials \citep{cifK9B8t}, within the text of the manuscripts (Figure \ref{fig:manubot-workflow}).
GitHub Actions runs a nightly workflow to update these external data and regenerate the statistics and figures for the manuscript.
The workflow uses the GitHub API to detect and save the latest commit of the external data sources that are GitHub repositories\footnote{Vaccines: \url{https://github.com/owid/covid-19-data}; Clinical Trials: \url{https://github.com/ebmdatalab/covid\_trials\_tracker-covid}; Cases and Deaths: \url{https://github.com/CSSEGISandData/COVID-19}}.
It then downloads versioned data from that snapshot of the external repositories and runs bash and Python scripts to calculate the desired statistics and produce the summary figures using Matplotlib \citep{1026Gxdsi}.
The statistics are stored in JSON files that are accessed by Manubot to populate the values of placeholder template variables dynamically every time the manuscript is built.
For instance, the template variable \texttt{\{\{ebm\_trials\_results\}\}} in the manuscript is replaced by the actual number of clinical trials with results, 98.
The template variables also include versioned URLs to the dynamically updated figures.
The JSON files and figures are stored in the \texttt{external-resources} branch of the GitHub repository, providing versioned storage.
The GitHub Actions workflow automatically adds and commits the new JSON files and figures to the \texttt{external-resources} branch every time it runs, and Manubot uses the latest version of these resources when it builds the manuscript.
The GitHub Actions workflow file is available online\footnote{\url{https://github.com/greenelab/covid19-review/blob/master/.github/workflows/update-external-resources.yaml}}, as are the scripts\footnote{\url{https://github.com/greenelab/covid19-review/tree/external-resources}}.
The Python package versions are also available\footnote{\url{https://github.com/greenelab/covid19-review/blob/external-resources/environment.yml}}.

Another issue identified was the need for standardized citation to clinical trials.
Other researchers identified the same need\footnote{\url{https://forums.zotero.org/discussion/74933/import-from-clinical-trials-registry} and \url{https://forums.zotero.org/discussion/77721/add-reference-from-clinical-trials-org}}.
Trials that are registered with clinicaltrials.gov receive a unique clinical trial identifier, or ``NCT ID.''
Because clinical trials are registered long before results are published, referencing clinical trial identifiers was a priority.
Manubot uses the Zotero translation server\footnote{\url{https://www.zotero.org} and \url{https://github.com/zotero/translation-server}} to extract citation metadata for some types of citations.
However, Zotero did not support clinical trial identifiers and could not extract relevant metadata from their URLs.
In order to pull clinical trial metadata associated into Manubot, we added Zotero support for these identifiers.
To achieve this, we query clinicaltrials.gov to retrieve XML metadata associated with each identifier using JavaScript\footnote{\url{https://github.com/zotero/translators/pull/2153}}.
This extension enables citing a trial as \texttt{@clinicaltrials:NCT04280705} instead of the URL.
Then, when Manubot requests clinical trial metadata from the Zotero translation server, the response includes the trial sponsors, responsible investigators, title, and summary.
Manubot now supports directly citing hundreds of registered Compact Uniform Resource Identifiers\footnote{\url{https://identifiers.org}}, beyond just the \texttt{clinicaltrials} identifier.

Because of the large number of citations used in this manuscript and the fast-moving nature of COVID-19 research, keeping track of retractions, corrections, and notices of concern also became a challenge.
We implemented a new Manubot plugin to support ``smart citations'' in the HTML build of manuscripts.
The plugin uses the scite \citep{14UJbLWf4} service to display a badge below any citation with a DOI.
The badge contains a set of icons and numbers that indicate how many times that source has been mentioned, supported, or disputed and whether there have been any important editorial notices.
We were thus able to identify references that needed to be reevaluated by an expert.
This addition was invaluable given the nature of the project, where we were disseminating rapidly evolving information of great consequence from over 1,500 different sources.
The badges also allow readers to ascertain a rough approximation of the reliability of cited sources at a glance.

Because most collaborators were writing and editing text through the GitHub website rather than in a local text editor, we also needed to add spell-checking functionalities to Manubot.
We integrated an existing Pandoc\footnote{\url{https://pandoc.org}} spell-check extension with AppVeyor CI to automatically post spelling errors as comments in a GitHub pull request.
The comment reported both unique misspelled tokens and all locations where the token was detected.
Project maintainers managed a custom dictionary to allow over 1,500 scientific and technical terms that were not common English words.
Spell-checking also helped standardize the writing style across dozens of authors by detecting features such as British versus American English spellings.
The actual spell-checking was implemented using GNU Aspell\footnote{\url{http://aspell.net}} and the Pandoc spellcheck filter\footnote{\url{https://github.com/pandoc/lua-filters/tree/master/spellcheck}}.
The filter enables checking only the manuscript text, ignoring URLs and formatting.

Manubot can render a manuscript in several formats that serve different purposes.
Prior to this project, Manubot could use Pandoc to convert the markdown-formatted manuscript to HTML, PDF, and DOCX formats.
We expanded this functionality to export individual sections of the manuscript as separate DOCX files while still rendering the complete manuscript in HTML and PDF formats.
This development was necessary because the manuscript grew so large that it needed to be split into seven separate papers for journal submission while still maintaining shared GitHub discussion across topics.
When exporting an individual section, Manubot customizes the manuscript title, authors, and author contributions to pertain to that specific section.
In addition, we expanded the export formats to include partial LaTeX support via Pandoc.
Pandoc converts the markdown content for an individual section to TeX and the Citation Style Language JSON, which contains reference metadata generated by Manubot, to BibTeX.
We customized a LaTeX template and reformatted the Manubot metadata, such as authors and their affiliations, for the LaTeX template.
The exported TeX file requires manual refinement but contains all manuscript content and most of the formatting.
Because LaTeX is required for manuscript submission in many fields, automating most of the process of converting markdown to a submission-friendly format expands Manubot's potential user base.
Manubot users can write in the simple markdown format, render the manuscript in continuously-updated PDF or interactive HTML formats, and export the manuscript in DOCX or TeX and BibTeX for submission to traditional publishers, taking full advantage of Pandoc's powerful document conversion capabilities and Manubot's automation.

\hypertarget{results}{%
\section{RESULTS}\label{results}}

\hypertarget{recruitment-and-manuscript-development}{%
\subsection{Recruitment and Manuscript Development}\label{recruitment-and-manuscript-development}}

\begin{figure}[htbp]
\hypertarget{fig:projectstats}{%
\centering
\includegraphics[width=0.5\textwidth]{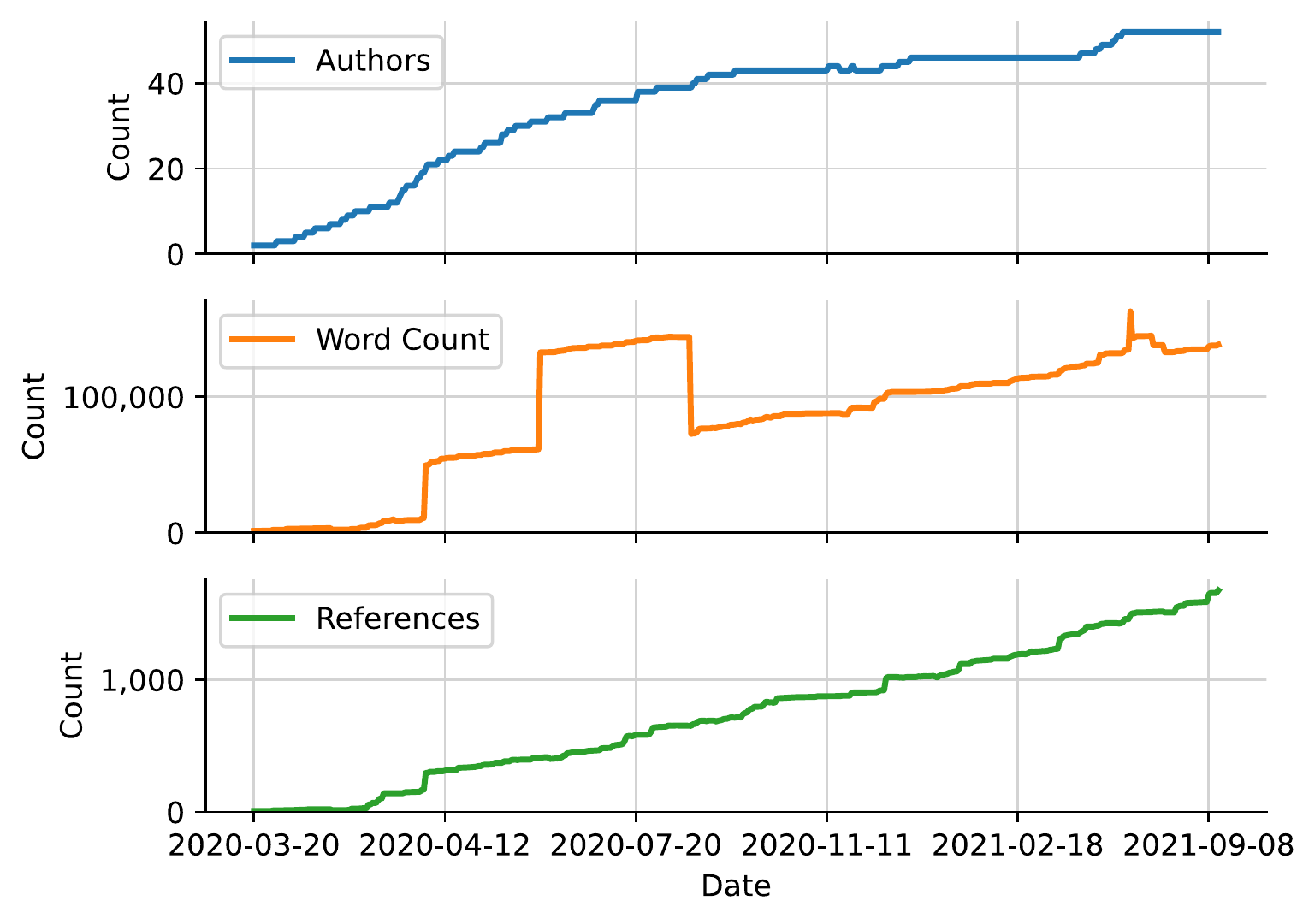}
\caption{\textbf{Project growth over time.}
The number of authors, word count, and number of references have all grown dramatically from when the project began on March 20, 2020.
As of September 10, 2021, there were 52 authors (including consortia), 1,676 references, and 138,213 words.
The spike in word count during summer 2020 was caused by erroneous duplication and subsequent removal of a large appendix.}\label{fig:projectstats}
}
\end{figure}

\begin{figure}[htbp]
\hypertarget{fig:projectdots}{%
\centering
\includegraphics[width=0.5\textwidth]{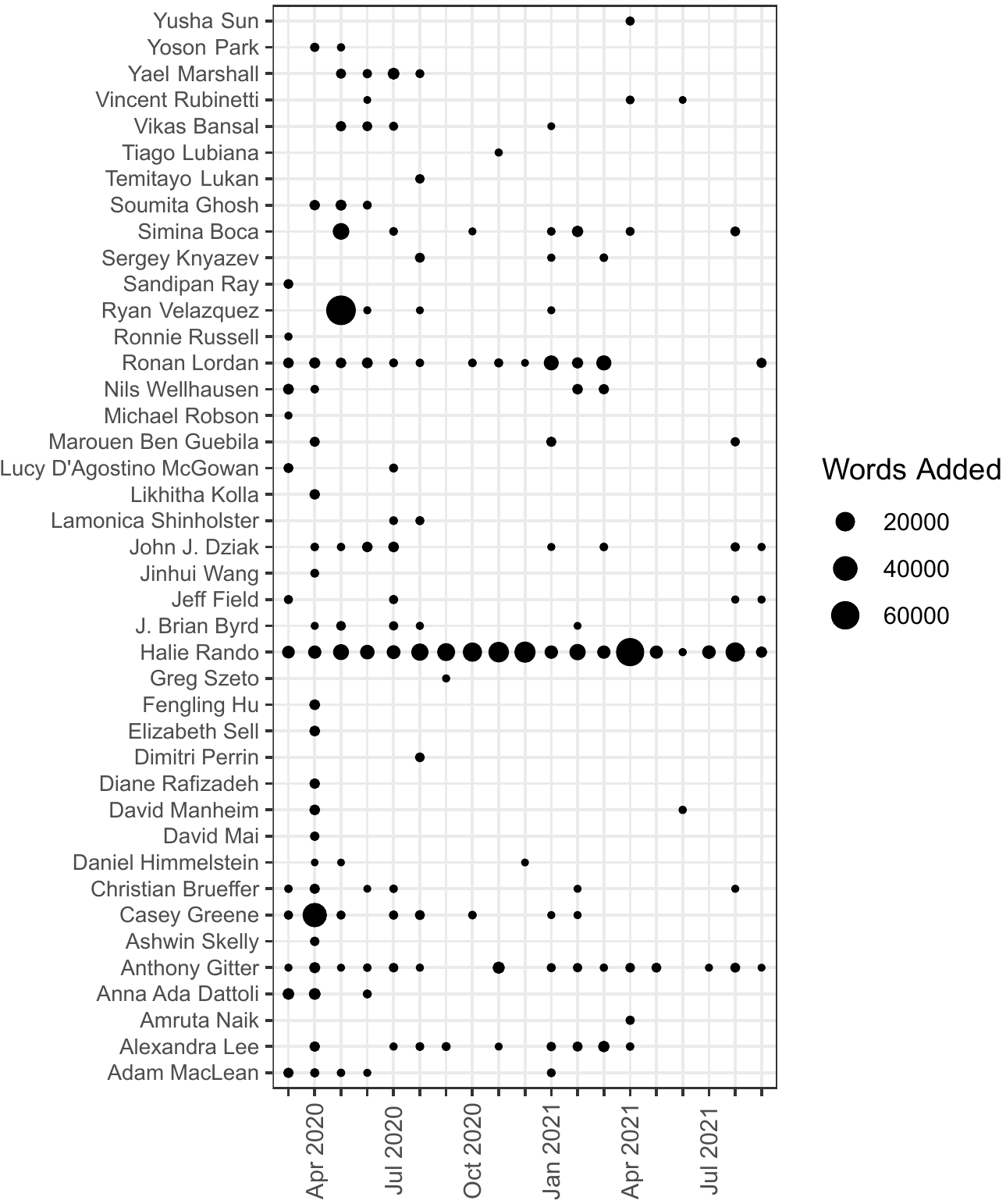}
\caption{\textbf{User contributions to the manuscript text over time.}
The dot size indicates the number of words added or edited each month since March 2020.
The figure does not depict other types of author contributions such as literature summaries, pull request review, visualization, or software.}\label{fig:projectdots}
}
\end{figure}

Coverage by \emph{Nature Toolbox} \citep{AE0QcVgJ} and an associated tweet\footnote{\url{https://twitter.com/j\_perkel/status/1245454628235309057}} about the project on April 1, 2020 attracted the interest of the scientific community (Figure \ref{fig:projectstats}).
Because GitHub issues are similar to other common web commenting systems, authors learned these tools quickly.
The Gitter chat also presented a low barrier to entry.
The manuscript continued to grow throughout the first year and a half of the project in both word count and the number of references (Figure \ref{fig:projectstats}).
Though only a fraction of potential contributors contributed to the text included in the manuscripts (Figure \ref{fig:projectstats}), many contributors remained engaged over the long term (Figure \ref{fig:projectdots}).
Additionally, new contributors continued to join even into the second year of the project.

In order to make the project more accessible, we developed resources explaining how to use GitHub's web interface to develop and edit text for Manubot assuming no prior experience with version control.
These tutorials explained how to open an issue, open a pull request, and review a pull request\footnote{CONTRIBUTING.md and INSTRUCTIONS.md within the repository}.
Additionally, the framework for evaluating literature was converted into issue templates to simplify the review of new articles.
Articles were classified as \emph{diagnostic}, \emph{therapeutic}, or \emph{other}, with an associated template developed to guide the review of papers and preprints in each category.
A total of 285 new paper issues had been opened as of September 13, 2021.

The manuscripts produced by the consortium (excluding this one) will be submitted to \emph{mSystems} as part of a special issue that provides support for continuous updates as more information becomes available.
One has been published and two are available as preprints.
This approach allows for a version of record to be maintained alongside the most recent version, which is always available through GitHub.
These manuscripts cover a wide range of topics including the fundamental biology of SARS-CoV-2 (pathogenesis \citep{r366f5T3} and evolution), biomedical advances in responding to the virus and COVID-19 (pharmaceuticals \citep{cifK9B8t}, nutraceuticals \citep{wgAGKcBj}, vaccines, and diagnostic technologies), and biological and social factors influencing disease transmission and outcomes.
To date, 50 authors are associated with the consortium (Figure \ref{fig:projectstats}).

More formal recruitment efforts to integrate with existing projects providing support for undergraduate students during COVID-19 were also successful.
We incorporated summaries written by the students, post-docs, and faculty of the Immunology Institute at the Mount Sinai School of Medicine\footnote{\url{https://github.com/ismms-himc/covid-19\_sinai\_reviews}} \citep{YZ4cHNuH}.
Additionally, two of the consortium authors were undergraduate students recruited through the American Physician Scientist Association's Virtual Summer Research Program.
Thus, the consortium was successful in providing a venue for researchers across all career stages to continue investigating and publishing at a time when many biomedical researchers were unable to access their laboratory facilities.

\hypertarget{integrating-data}{%
\subsection{Integrating Data}\label{integrating-data}}

We integrated data into the manuscripts from several sources (Figure \ref{fig:manubot-workflow}).
Worldwide cases and deaths were tracked by the COVID-19 Data Repository by the Center for Systems Science and Engineering at Johns Hopkins University\footnote{\url{https://github.com/CSSEGISandData/COVID-19/tree/master/csse\_covid\_19\_data/csse\_covid\_19\_time\_series}}.
The clinical trials statistics and figure were generated based on data from the University of Oxford Evidence-Based Medicine Data Lab's COVID-19 TrialsTracker \citep{SSbnPnzT}.
Information about vaccine distribution was extracted from Our World In Data\footnote{\url{https://github.com/owid/covid-19-data}} \citep{ZHvhFakW}.
Figure \ref{fig:cord19-growth} integrates data from the CORD-19 dataset \citep{CiOwklc6}.

Manubot's bibliographic management capabilities were critical because the amount of relevant literature published far outstripped what we had anticipated at the beginning of the project.
As of September 10, 2021, there were 1,676 references (Figure \ref{fig:projectstats}).
The scite plugin provided a way to visually inspect the reference list to identify possible references of concern.
This and the other new features required for the COVID-19 project are now included in Manubot's rootstock, which is the template GitHub repository for creating a new manuscript.
Using CI, Manubot now checks that the manuscript was built correctly, runs spell-checking, and cross-references the manuscripts cited in this review.
In addition, Manubot now supports citing clinical trial identifiers such as \texttt{clinicaltrials:NCT04292899} \citep{yTCAmOyt}.

\hypertarget{discussion}{%
\section{DISCUSSION}\label{discussion}}

The current project was based in the GitHub repository \texttt{greenelab/covid19-review} using Manubot \citep{YuJbg3zO} to continuously generate the manuscript.
The Manubot framework facilitated a massive collaborative review on an urgent topic.
We demonstrated the utility of Manubot to a project where many contributors lacked expertise or even experience working with version control.
This effort has produced not only seven literature reviews on topics relevant to the COVID-19 pandemic, but has also generated cyberinfrastructure for training novice users in GitHub.
We also extended the functionalities of Manubot to provide more of the benefits of What You See Is What You Get platforms such as Google Docs (Table \ref{tbl:manubot-addons}).
Open publishing thus allowed us to harness the domain expertise of a large group of non-technical users to respond to the flood of COVID-19 publications.

Several existing and new features in Manubot aid in responding to the challenges posed by the infodemic.
Manuscripts are written in markdown and can be rendered in several formats providing different advantages to users.
For example, beyond building just a PDF, Manubot also renders the manuscript in HTML, DOCX, and now, LaTeX (in a more limited capacity).
The interactive HTML manuscript format offers several advantages over a static PDF to harmonize available resources and address specific problems related to COVID-19.
The integration of scite into the HTML build makes references more manageable by visually indicating whether their results are contested or whether they have been corrected or retracted.
Cross-referencing different pieces of the manuscript, such as cited preprints with reviews stored in an appendix, is another interactive option presented by HTML.
The DOCX format was preferred by most non-technical users for reviewing the final version of the manuscript and was useful for creating submissions to a biological journal.
Additionally, because of the heavy emphasis on Word processing in biology, Manubot's ability to generate DOCX outputs was expanded to allow users to generate DOCX files containing only a section of the manuscript.
In our case, where the full project is nearly 150,000 words, this allows individual pieces to be shared more easily.
Finally, the preliminary addition of LaTeX output is useful for researchers from computational fields who submit papers in TeX format and removes the step of reformatting markdown prior to submission.

\begin{table}[ht]
  \caption{\textbf{Manubot extensions for the COVID-19 review.}}
  \begin{tabular}{p{0.06\textwidth} p{0.38\textwidth}}
    \toprule
    \textbf{Type} & \textbf{Description} \\
    \midrule
    CI & Regularly download external data sources, generate new figures and statistics, and read them when Manubot builds the latest manuscript \\
    CI & Post spell-checking reports as pull request comments \\
    Citations & Zotero extension to report more relevant clinical trial metadata from \url{https://clinicaltrials.gov} \\
    Citations & Cite any Compact Uniform Resource Identifier, such as \texttt{clinicaltrials} or \texttt{ncbigene} \\
    Citations & scite badges to track retractions, corrections, and notices of concern \\
    Outputs & Improved support for Pandoc's LaTeX output \\
    Outputs &  Build complete manuscript alongside individual sections as standalone documents \\
  \bottomrule
\end{tabular}
\label{tbl:manubot-addons}
\end{table}

The COVID-19 Review Consortium provided a platform for researchers to engage in scientific investigation early in the pandemic when many biological scientists were unable to access their research spaces.
In turn, by seeking to adapt Manubot to allow for broader participation, we made a number of improvements that are expected to increase its appeal to researchers from all backgrounds.
Manubot provided a way for contributors from a variety of backgrounds, including early-career researchers, to join a massive collaborative project while demonstrating their individual contributions to the larger work and gaining experience with version control.
The licensing and infrastructure also provide the basis for individuals to adapt from this project to create their own snapshots of the COVID-19 literature that derive from, but are not wholly identical to, the primary versions of these reviews.
This project suggests that massive online open publishing efforts can indeed advance scholarship through inclusion \citep{PoDz2q0A}, including during the extreme challenges presented by the COVID-19 pandemic.

Some challenges did arise in efforts to include an academically diverse set of authors.
The barriers to entry posed by git and GitHub likely still reduced participation from individuals who might have otherwise been interested.
Using pull requests as a tool for writing text is also unfamiliar to many or most scientists, and the review process can be slow, which might cause interested contributors to lose interest.
Additionally, the pull request model may limit people from providing general feedback on the manuscript or a section of the manuscript.
As a result, some feedback came through email or comments on the DOCX outputs that were then translated into issues or pull requests by the project managers.
Given that our approach hinged on these version control tools, it is likely that our group of contributors was biased towards those who were interested in or experienced with computational tools.
The trajectory of the pandemic itself also likely influenced participation: engagement waned over the course of the pandemic as labs opened back up and researchers were able to return to their work, and we recruited very few senior clinicians to the project, which is unsurprising given the load on medical professionals during this time.
Engagement that waxes and wanes is, however, typical when writing massively open online papers \citep{PoDz2q0A}.
Adding features such as spell-check did improve usability, and additional features such as automatically checking the formatting of citations could further improve the usability of this tool.
In the future, a formal study of participation could allow for quantification of these biases and improved efforts to foster inclusion.

Additional limitations are challenges associated with massively open online papers in general.
With such a large amount of text, it is not possible to keep all sections of the manuscript up to date at all times.
Readers are not able to distinguish when each section was updated.
Even GitHub's blame functionality does not distinguish minor changes from substantive updates to the text.
While much of the data and statistics update automatically, the text itself required updating by human experts.
This asynchronicity could potentially introduce incompatibility between the figures and the surrounding text.
Similarly, in line with the collaboration-related challenges of the project, some authors returned to update their text, while others did not.
As a result, the lead authors of each paper often spent several weeks prior to journal submission updating the text to reflect new developments in each area.
In the future, it may be possible to streamline this process through integration with a tool such as CoronaCentral \citep{Ybg667S0} to automatically identify relevant, high-impact papers that need to be included, although expertise would still be required to incorporate them.
Another challenge involves tracking preprints as they are reviewed or critiqued, revised, and potentially published.
While updating the content of the manuscript would likely fall to human contributors, automatic detection of published versions of preprints \citep{pESBLU4c} could be integrated in the future.
These challenges are exacerbated by the scale of the infodemic, but developing solutions would benefit future projects tracking more typical trends in publication.
Similarly, outputting machine readable summaries of key information in the COVID-19 review manuscripts could reduce their contribution to the infodemic.
As it stands, the integration of Compact Uniform Resource Identifier does make a step in this direction.
Formal identifiers could be used to extract relationships among clinical trials, genes, publications, and other entities.
Thus, the experience of using Manubot for a massive project has laid the foundation for future additions to enhance user experience and inclusivity.

\hypertarget{conclusion}{%
\section{CONCLUSION}\label{conclusion}}

With the worldwide scientific community uniting during 2020 and 2021 to investigate COVID-19 from a wide range of perspectives, findings from many disciplines are relevant on a rapid timescale to a broad scientific audience.
As many other efforts have described, the publishing rate of formal manuscripts and preprints about COVID-19 has been unprecedented \citep{7ub6VM4Z}, and efforts to review the body of COVID-19 literature are faced with an ever-expanding corpus to evaluate.
In the case of the seven manuscripts produced by the COVID-19 Review Consortium, Manubot allows for continuous updating of the manuscripts as the pandemic enters its second year and the landscape shifts with the emergence of promising therapeutics and vaccines \citep{cifK9B8t}.
These manuscripts pull data from external sources and update information and visualizations daily using CI.
By off-loading some updates to computational pipelines, domain experts can focus on the broader implications of new information as it emerges.
Centralizing, summarizing, and critiquing data and literature broadly relevant to COVID-19 can expedite the interdisciplinary scientific process that is currently happening at an advanced pace.
As of September 13, 2021, 2,886 commits have been made to the manuscript across 575 merged pull requests.
The efforts of the COVID-19 Review Consortium illustrate the value of including open source tools, including those focused on open publishing, in these efforts.
By facilitating the versioning of text, such platforms also allow for documentation of the evolution of thought in an evolving area and formal analysis of a collaborative project.
This application of version control holds the potential to improve scientific publishing in a range of disciplines, including those outside of traditional computational fields.
While Manubot is a technologically complex tool, this project demonstrates that it can be applied to a variety of projects.
Future work can address remaining limitations and continue to advance Manubot as an inclusive tool for open publishing projects.

\hypertarget{acknowledgements}{%
\section*{Acknowledgements}\label{acknowledgements}}
\addcontentsline{toc}{section}{Acknowledgements}

We thank Josh Nicholson and Milo Mordaunt for scite support, David Nicholson for spell-check assistance, Milton Pividori and consortium members\footnote{COVID-19 Review Consortium: Vikas Bansal, John P. Barton, Simina M. Boca, Joel D Boerckel, Christian Brueffer, James Brian Byrd, Stephen Capone, Shikta Das, Anna Ada Dattoli, John J. Dziak, Jeffrey M. Field, Soumita Ghosh, Anthony Gitter, Rishi Raj Goel, Casey S. Greene, Marouen Ben Guebila, Daniel S. Himmelstein, Fengling Hu, Nafisa M. Jadavji, Jeremy P. Kamil, Sergey Knyazev, Likhitha Kolla, Alexandra J. Lee, Ronan Lordan, Tiago Lubiana, Temitayo Lukan, Adam L. MacLean, David Mai, Serghei Mangul, David Manheim, Lucy D'Agostino McGowan, Amruta Naik, YoSon Park, Dimitri Perrin, Yanjun Qi, Diane N. Rafizadeh, Bharath Ramsundar, Halie M. Rando, Sandipan Ray, Michael P. Robson, Vincent Rubinetti, Elizabeth Sell, Lamonica Shinholster, Ashwin N. Skelly, Yuchen Sun, Yusha Sun, Gregory L Szeto, Ryan Velazquez, Jinhui Wang, Nils Wellhausen} Alex Lee and Christian Brueffer for feedback, and Nick DeVito for assistance with the Evidence-Based Medicine Data Lab COVID-19 TrialsTracker data.
Research was supported by the Gordon and Betty Moore Foundation award GBMF 4552 (HMR, DSH, CSG), the National Institutes of Health award R01HG010067 (HMR, CSG), and the John W. and Jeanne M. Rowe Center for Research in Virology (AG)\footnote{Conflicts of interest. SMB: Now employed by AstraZeneca (Gaithersburg, MD). May own stock or stock options. Work conducted at previous position. LDM: Received consulting fees from Acelity and Sanofi. AG: Patent application filed with the Wisconsin Alumni Research Foundation related to classifying activated T cells.}.

\bibliography{methods.bib}

\end{document}